\documentclass[11pt]{article}

\usepackage[final]{acl}

\usepackage{times}
\usepackage{latexsym}

\usepackage[T1]{fontenc}

\usepackage[utf8]{inputenc}

\usepackage{microtype}

\usepackage{inconsolata}

\usepackage{soul}
\usepackage{multirow}
\usepackage{graphicx}
\usepackage{amsmath}
\usepackage{xcolor}
\usepackage{etoolbox}
\usepackage{todonotes}

\title{BLADE: Better Language Answers through Dialogue and Explanations}

\author{Chathuri Jayaweera, Phoebe Huang, Bonnie J. Dorr \\ University of Florida, Gainesville, FL, USA\\
  \texttt{\{chathuri.jayawee, pg.huang, bonniejdorr\}@ufl.edu} \\}

\begin{document}
\maketitle
\begin{abstract}
Large language model (LLM)-based educational assistants often provide direct answers offering little incentive for students to explore or engage with course materials. We present BLADE (Better Language Answers through Dialogue and Explanations), a retrieval-augmented generation (RAG) based conversational assistant grounded in course-specific content that guides students toward relevant materials through citation-grounded dialogue rather than delivering unsourced solutions. We evaluate BLADE in an advanced undergraduate NLP course with extensive instructional resources, in which locating and synthesizing relevant material is a central challenge. During quizzes, students are assigned to one of three conditions: BLADE only, direct course materials only, or both; we measure performance and resource usage. Results show that students consistently select BLADE over direct materials when both are available, and that quiz performance is highest when students use BLADE alone. In contrast, simultaneous use of both resources is associated with significantly lower performance. No statistically significant difference in quiz performance occurred between assigned resource conditions. These findings show the potential of RAG-based assistants as a citation-grounded interface for applying course content under open-book exam-like conditions, while suggesting that adding direct-material access does not necessarily improve in-task performance.

\end{abstract}

\section{Introduction}
In advanced, resource-intensive courses such as Natural Language Processing (NLP), the rapid expansion of digital learning materials creates a paradox: while students have unprecedented access to instructional content, they often struggle to identify the content that bridges a question to a coherent conceptual understanding. Textbooks, lecture slides, recorded lectures, and curated readings are typically distributed across multiple platforms, leaving learners to manually search for and synthesize relevant information.

Recent advances in large language models (LLMs) have driven widespread adoption of conversational assistants for educational support. Most existing systems are developed and evaluated in constrained instructional settings (e.g., introductory programming tasks), where content scope and required background knowledge are limited \cite{kucharavy_llms_2025}. In contrast, advanced undergraduate courses like NLP involve a broader, more interconnected set of concepts, requiring students to navigate and integrate information across a large body of material while still engaging in independent problem solving.

Existing systems typically prioritize direct answer generation, offering limited opportunities for exploration, self-explanation, and sustained engagement with instructional resources, as reflected in recent BEA shared tasks that frame tutoring as teacher response generation in dialogue settings \cite{kochmar_proceedings_2023}. Prior work in learning sciences
emphasizes active inquiry, evidence-based reasoning, and interaction with authoritative
materials for durable conceptual understanding \cite{sciences_how_2018, jong_lets_2023}. 

Recent work in educational NLP emphasizes interactive guidance, mistake remediation, and pedagogically grounded feedback rather than simply providing correct answers \cite{kochmar_findings_2025}. At the same time, overreliance on AI-generated responses raises concerns about passive learning, misplaced trust, and reduced verification behavior in the presence of hallucinated or oversimplified outputs \cite{kucharavy_llms_2025, beale_dialogic_2025}.

\begin{figure}[h]
\small
\centering
\includegraphics[scale=0.27]{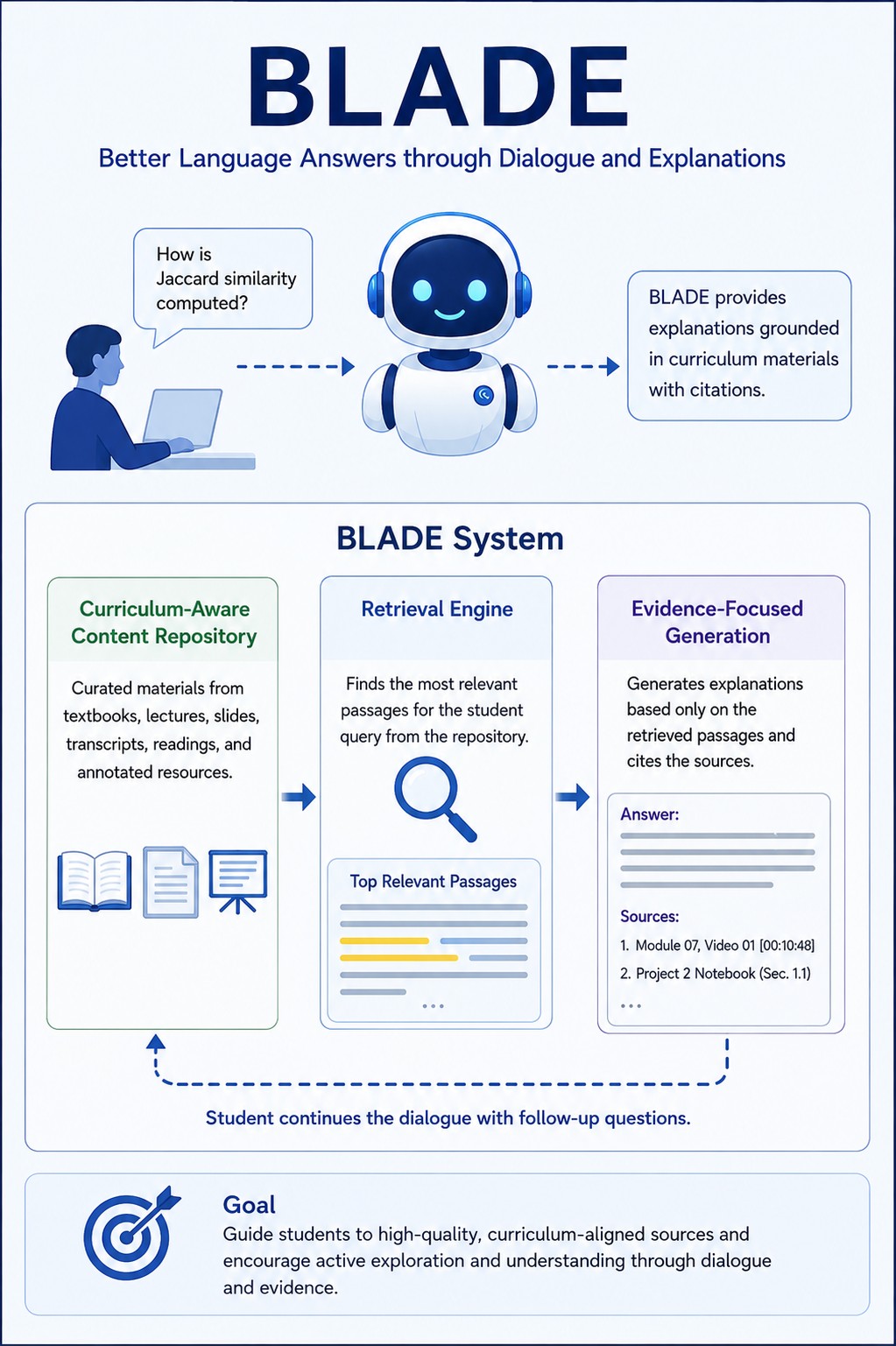}
\vspace*{-.05in}
\caption{BLADE surfaces relevant instructional resources for student queries. BLADE directs students to supporting evidence, fostering explanation-based learning and active exploration.}
\vspace{-0.27in}
\label{Fig:BLADE-Diagram}
\end{figure}

In this work, we introduce BLADE (Better Language Answers through Dialogue and Explanations), a conversational assistant implemented and empirically evaluated in a high-resource instructional setting. BLADE is designed to guide learners toward relevant instructional resources through anchoring its responses in retrieved, pedagogically relevant course content, including textbooks, lecture slides, readings, and annotated resources. It presents citations to source material, so that explanations remain traceable to authoritative passages rather than presented as standalone answers (Figure~\ref{Fig:BLADE-Diagram}). Responses may take the form of structured explanations (examples are provided in Appendix~\ref{appendix:queries-and-responses}), but each is anchored to, and cites, specific instructional sources, encouraging students to consult them directly.

Our contributions are: (1) a retrieval-grounded instructional assistant that guides students to authoritative, curriculum-aligned passages and uses them to ground conversational interaction; (2) a prompt-based approach to retrieval-centered assistance in resource-rich learning 
environments, implemented through system-prompt configuration of an existing RAG framework to provide structured access to instructional materials and encourage exploration;
and (3) a classroom impact study evaluating BLADE in an undergraduate computer science course, where student resource usage and post-instructional quiz performance are analyzed across three conditions: BLADE alone, direct course materials alone, and both combined. This design allows us to examine quiz performance across resource conditions and how students choose to use available resources.

By anchoring each exchange in vetted instructional content, BLADE provides transparent evidence for its guidance, reducing the risk of misinformation and hallucination. This design encourages active learning behaviors (e.g., elaboration, self-explanation, and retrieval practice) by prompting learners to engage directly with source material, which prior work has shown to support long-term retention and transfer. BLADE combines the natural fluency of dialogue with verifiable, traceable references, providing
transparent access to instructional support.


We present the design, implementation, and evaluation of BLADE in an advanced undergraduate NLP course with extensive 
resources. Topics include text preprocessing, corpus analysis, CKY parsing, document similarity, hidden Markov models, part-of-speech tagging, semantic role labeling, machine translation, encoder-decoder transformers, and model evaluation. Quizzes, exams, and programming projects require students to apply these concepts in both theoretical and practical contexts.

Our results show that when both are available, students 
prefer BLADE over direct course materials. Across resource configurations, BLADE-only use is associated with the highest average quiz scores, although its advantage over other single-resource or no-resource conditions is not statistically significant. Combined-resource use is associated with significantly lower scores. This pattern may reflect the cognitive demands of combining resources or a tendency for students struggling with the material to seek additional support.


These findings highlight the potential of structured retrieval-based conversational assistants as standalone learning tools and underscore the importance of careful instructional design when integrating AI-assisted and direct-access resources. Our conclusions apply only to the resource conditions evaluated in Section~\ref{sec:classroom-deployment} and not to assistants based on other paradigms, such as direct answer generation. Determining how answer-generating and exploration-oriented assistants differ in their effects on student learning requires a direct comparison beyond the scope of this study.



\section{Related Work}


Retrieval-augmented generation (RAG) is central to educational AI, combining information retrieval with the expressive capabilities of large language models.
Grounding responses in retrieved materials reduces hallucinations compared with purely generative approaches \cite{lewis_retrieval-augmented_2020, ji_survey_2023, gao_retrieval-augmented_2024}, while retrieval of textbooks and lecture notes improves factual accuracy and provides unified access to course knowledge 
\cite{2024.EDM-short-papers.28, li_retrieval-augmented_2025, lee_developing_2025}. 

RAG-based tutors improve student knowledge retention over unassisted baselines, though they may still lag behind stronger generative models  \cite{slade_transforming_2024}. Because they are typically evaluated without direct student access to the underlying materials, it remains unclear what the assistant contributes beyond the resources themselves, particularly in high-resource courses.

A parallel line of work explores tutoring agents that promote engagement by withholding direct answers. Step-based dialogue tutors can approach the effectiveness of human one-on-one instruction \cite{vanlehn_relative_2011}, motivating LLM-based systems grounded in Socratic questioning and scaffolded hint generation \cite{hutchison_automating_2014, alshaikh_socratic_2020, favero_enhancing_2024, liu_socraticlm_2024, ding_boosting_2024}. These approaches improve understanding and reflective thinking in higher-education settings \cite{xi_investigating_2025}, although poorly aligned guidance can introduce cognitive overload and hinder learning \cite{liu_discerning_2025}. Despite emphasizing engagement, these systems still center learning on AI-generated responses and are rarely evaluated in terms of how students interact with the underlying instructional materials.



Recent work combines retrieval grounding with guided interaction \cite{li_retrieval-augmented_2025, dong_how_2025, modran_llm_2024}, including knowledge-graph-enhanced retrieval \cite{dong_how_2025}. Educational studies highlight trade-offs between fidelity to source materials and student preferences \cite{2024.EDM-short-papers.28}, suggesting that retrieval quality alone does not ensure pedagogical effectiveness. Yet these systems continue to emphasize answer generation, making it difficult to separate the contribution of the assistant from that of the retrieved resources, particularly when both AI-mediated and direct resource access are available.



Educational theory supports guiding students to primary 
resources rather than providing answers. Constructivist theory \cite{piaget_understand_1980} and retrieval-practice research  \cite{roediger_critical_2011} show that locating and interpreting information strengthens memory and understanding. Dialogic pedagogy cautions that direct answers can undermine knowledge co-construction \cite{beale_dialogic_2025}. In computer science education, guided discovery and scaffolding improve problem-solving transfer through engagement with instructional materials \cite{kelleher_storytelling_2007,kestin_ai_2025}. These findings motivate designs that guide students to relevant materials while preserving independent reasoning.

Recent work has introduced pedagogical evaluation frameworks for LLM-powered tutors (e.g., \citet{maurya-etal-2025-unifying, macina-etal-2025-mathtutorbench}), offering valuable intrinsic assessments of instructional quality. While these complement the outcome-based evaluation presented here, our primary goal is to examine BLADE's relationship to quiz performance and resource engagement in an authentic classroom setting. Integrating intrinsic metrics with classroom outcomes remains an open direction for future research.


Taken together, prior work highlights three gaps. First, RAG-based tutoring systems rarely isolate the assistant's contribution from the instructional materials it retrieves \cite{li_retrieval-augmented_2025, slade_transforming_2024}. Second, retrieval is typically optimized for topical similarity rather than pedagogical relevance \cite{2024.EDM-short-papers.28, dong_how_2025}. Third, little is known about how students balance AI-assisted and direct resource use when both are available, particularly in high-resource settings \cite{slade_transforming_2024, xi_investigating_2025}. We address these gaps by evaluating a design centered on structured access to instructional materials in an advanced, resource-intensive NLP course, rather than answer generation. Under controlled resource conditions, we measure 
resource usage and quiz performance.




\section{BLADE System Architecture}

BLADE is built on a retrieval-augmented generation (RAG) framework that integrates curriculum-specific content with a conversational LLM interface. The system consists of three primary components: (i) a curriculum-aware content repository, (ii) an instructional relevance retriever, and (iii) a controllable generation module designed to foreground evidence rather than direct answers.

\subsection{Curriculum-Aware Content Indexing}

Course materials---including textbooks, lecture slides, readings, and supplementary documents---are segmented into semantically coherent instructional units. These units are annotated with metadata such as topic alignment, learning objectives, and instructional context (e.g., lecture
or textbook chapter). This structured indexing enables fine-grained retrieval of pedagogically relevant content in response to learner queries.

\begin{figure}[ht]
    \small
    \centering
    \includegraphics[scale=0.26]{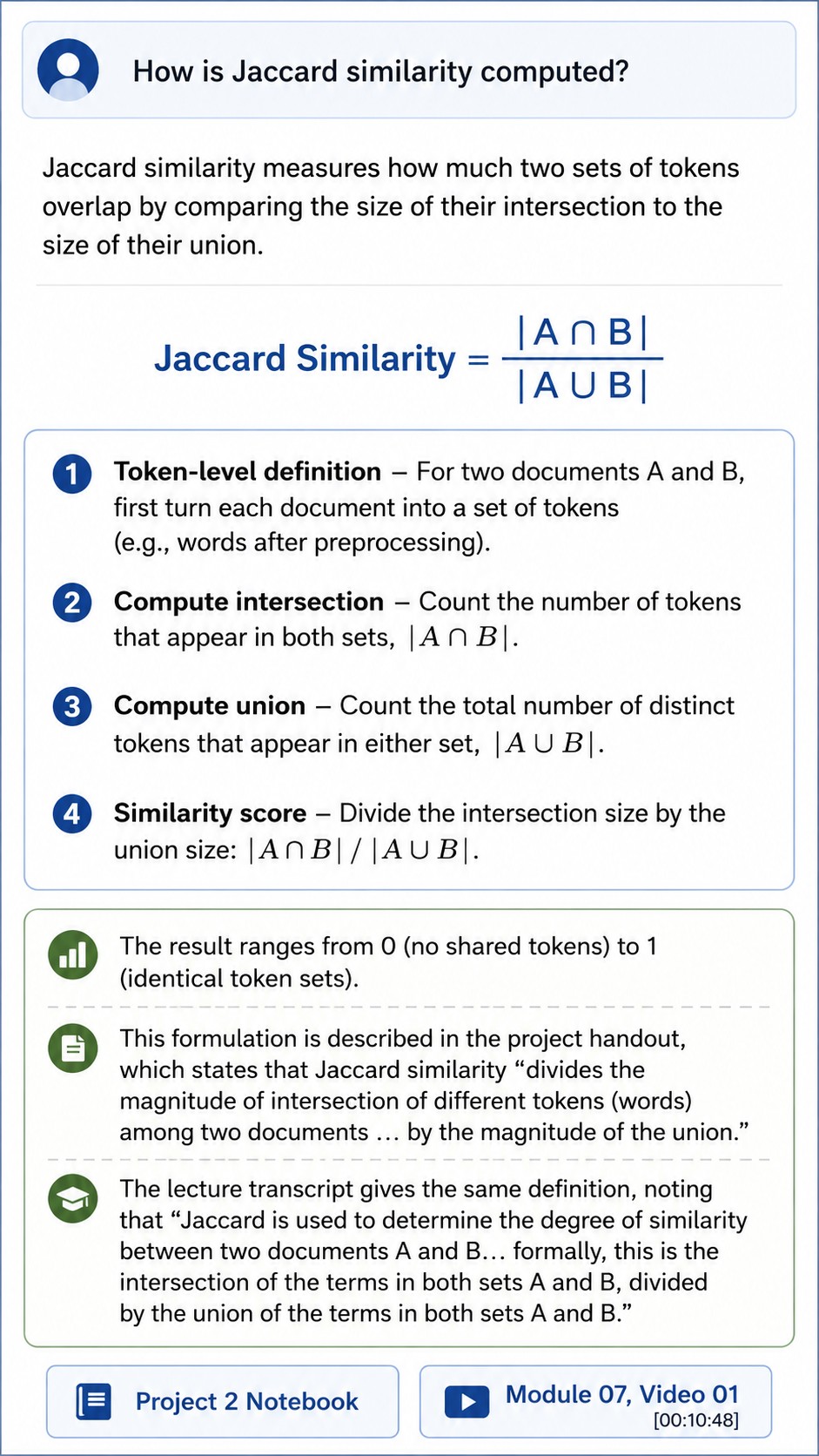}
    \vspace{-.1in}
    \caption{An example of a typical BLADE response to a query with citations of sources. The cited sources include textbook/course material excerpts and lecture transcripts with timestamps of the relevant portion.}
    \label{fig:blade-query-response}
    \vspace*{-.15in}
\end{figure}

\subsection{Instructional Relevance Retrieval}

BLADE uses hybrid retrieval over course materials, while its system prompt (Section~\ref{sec:evidence-focused-diag}) guides responses toward curriculum-aligned content. 
BLADE is implemented using an Onyx-based, institution-wide chat agent framework. Its retrieval pipeline incorporates Vespa hybrid search over course materials, with weighting parameter (alpha) set to 0.5. The system uses nomic-embed-text-v1.5\footnote{\url{https://huggingface.co/nomic-ai/nomic-embed-text-v1.5}} to generate normalized, 768-dimensional embeddings.
The retrieval component uses chunks of up to 512 tokens and initially retrieves 50 candidate documents. The top 20 candidates are selected for post-processing using an LLM-based relevance filter.
The selected passages provide instructional context for responses intended to support conceptual understanding and learning progression.


Figure~\ref{fig:blade-query-response} 
illustrates how BLADE grounds a student query about the Jaccard similarity in the retrieved project notebooks and lecture transcripts. Responses include citations and lecture timestamps to help students locate supporting content. Additional examples appear in
Appendix~\ref{appendix:queries-and-responses}.
\vspace{-.05in}

\subsection{Evidence-Focused Dialogue Generation} \label{sec:evidence-focused-diag}

BLADE is designed to support rather than replace student reasoning. Its responses are grounded in retrieved instructional materials and encourage students to explore supporting course content.

BLADE combines refusal behavior for unsupported or inappropriate requests with guidance grounded in retrieved course materials. It identifies out-of-scope requests and grounds permitted responses in course materials. For programming requests, BLADE declines to provide solutions but may offer course-grounded conceptual guidance
(see Appendix~\ref{appendix:queries-and-responses}). 

This behavior is reinforced through system-prompt constraints that define the instructional scope of the assistant. The default generator is gpt-oss-120b (117B parameters; \cite{openai_gpt-oss-120b_2025}). Prompt instructions emphasize retrieval-grounded guidance over direct problem-solving, including directives to rely on provided course materials, avoid solving assignments or providing code solutions, and redirect inappropriate requests toward course-relevant concepts. System-prompt constraints also are designed to reduce hallucinations by restricting BLADE to the retrieved course materials and requiring acknowledgment when sufficient supporting evidence is unavailable (no dummy citations). Examples are provided in Figure~\ref{fig:constraining-prompts}.

\begin{figure}[h]
\centering\small
\begin{tabular}{|p{2.8in}|}\hline
 --\textit{``Your role is to help students understand course concepts, find information in course materials, and clarify course logistics---nothing more''}\\
 --\textit{``Refer to...provided files, for course content questions''}\\
 --\textit{`Do not provide code solutions, homework answers, or solve problems for students''}\\
 --\textit{``Do your best: If the query seems relevant to the course, try to find an answer using both Module [lecture] transcripts and [textbook] chapters''}\\\
 --\textit{``If asked to do something inappropriate, decline politely and redirect to course-related questions''}\\ 
 --\textit{``ONLY use information provided in the context---do not use any external knowledge''}\\
 --\textit{``No dummy citations: If 
no relevant sources are found, admit that the knowledge base is insufficient to answer the question''}\\  \hline
\end{tabular}
    \caption{System-prompt constraints define BLADE's instructional scope and apply hallucination mitigation.}
    \label{fig:constraining-prompts}
    \vspace*{-.1in}
\end{figure}


BLADE cites the underlying course materials so that students can verify its responses. Although this grounding aims to improve factual reliability, evaluating response accuracy and pedagogical quality remains outside the scope of this work and is left for future research.

\section{Classroom Deployment and Evaluation}
\label{sec:classroom-deployment}

We evaluate BLADE in a Fall 2025 impact study conducted in an advanced undergraduate NLP course. All enrolled students are given access to BLADE from the start of the semester and interact with the assistant alongside standard course materials during homework assignments and study sessions. Of about 200 enrolled students, 85 undergraduate computer science students elect to participate, completing three quizzes mid-semester after recruitment via a study announcement and interest form (see Appendices~\ref{appendix:recruitment_form} and~\ref{appendix:quiz_instructions}).

\subsection{Study Context and Participants}


Participants are randomly divided into three groups and are assigned the same three NLP quizzes across three course modules. Before each quiz, participants are instructed to review the corresponding course materials, including PlayPosit videos with embedded thought questions. Each quiz contains an average of six automatically graded multiple-choice questions that assess conceptual understanding and application rather than extended derivations or programming.

Each group completes all three quizzes, with the available resource condition rotated across quizzes according to a fixed permutation, yielding nine quiz–resource combinations in total.
Each assignment is followed by a survey that records how students use their allowed resources. These data are used to evaluate students' resource choices,
student performance, and the relationship between performance and resource availability and usage.


BLADE is evaluated along three dimensions: (i) resource use when multiple modalities are available; (ii) performance across three resource conditions: BLADE alone, direct materials alone, and both combined; and (iii) the relationship between resource use and learning outcomes.


\subsection{Resource Configurations}

\begin{figure*}
\centering
\includegraphics[width=0.95\linewidth]{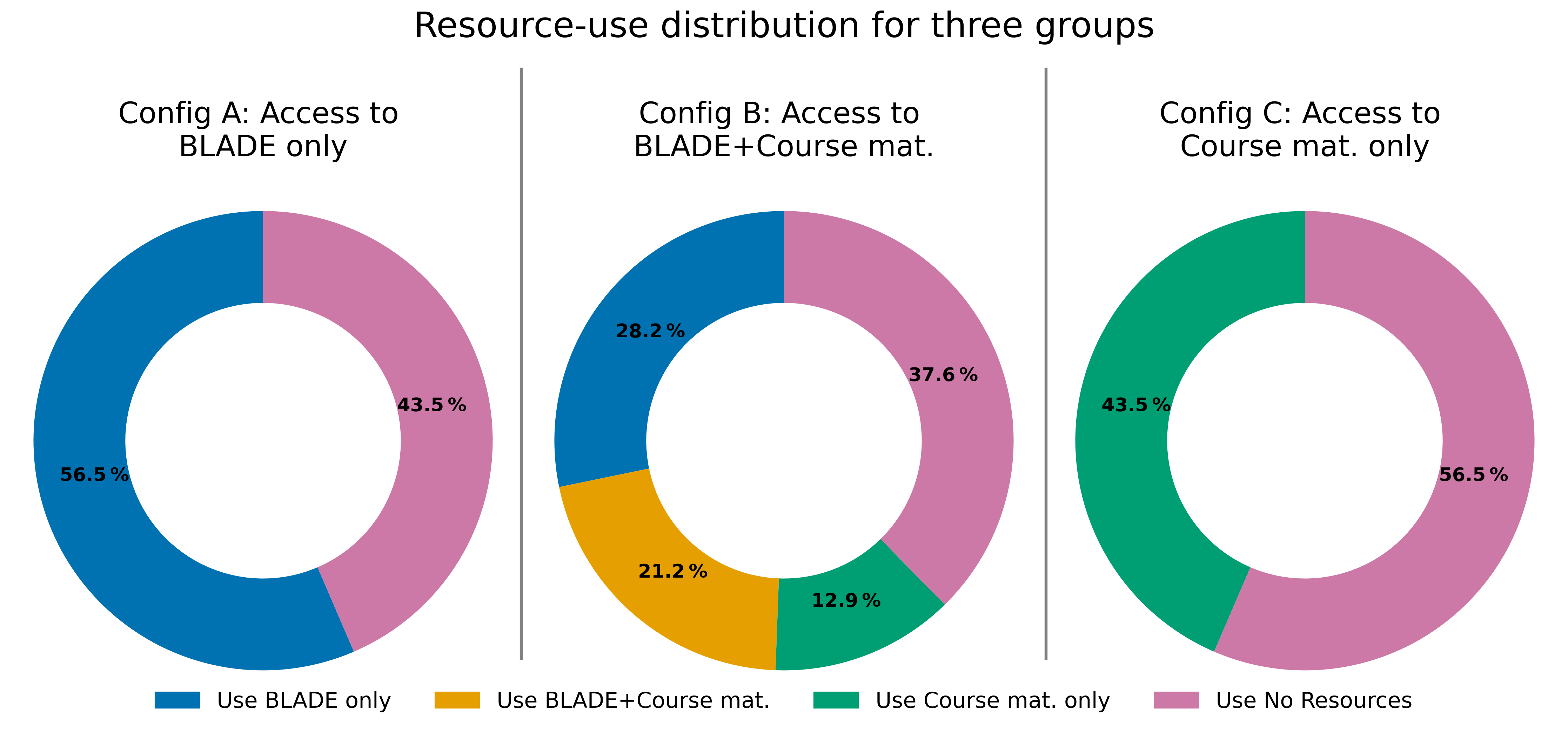}
\vspace*{-.15in}
\caption{
Resource use by availability condition: BLADE only (Config A: 
BLADE 48, none
37); BLADE + course materials (Config B: 
BLADE 
24, both 
18, course materials only
11, none 48); and 
course materials only (Config C:
course materials only 37, none 48).
Counts represent
mutually exclusive categories of actual resource use.
}
\vspace*{-0.15in}
\label{fig:survey-res-use}
\end{figure*}

Students complete quizzes in a controlled environment with access only to the resources assigned to their condition. Quizzes are administered under remote proctoring using Honorlock, which prevents access to browser windows or tabs beyond those permitted, thereby blocking unauthorized external tools, including general-purpose LLMs such as ChatGPT. Resource use is voluntary: students may use any, all, or none of the available resources (Figure~\ref{fig:survey-res-use}). For example, students in the BLADE-plus-materials condition may use only BLADE, only the course materials, both, or neither.

This design allows us to observe not only whether access to a resource improves performance, but also how students choose to engage with it when given the opportunity. Post-quiz surveys allow us to triangulate self-reported engagement with observed performance.

Together, the quiz results and usage data allow us to address two related questions: whether access to BLADE improves performance relative to direct materials alone, and whether students who engage with BLADE perform differently from those who, despite having access, do not use it.


\subsection{Group Assignments} \label{sec:group-assignment}

The quiz configurations are defined as follows:
\begin{tabular}{lp{2.7in}}
$\bullet$ & Config A: access to BLADE but no other course resources.\\
$\bullet$ & Config B: access to BLADE plus all course resources.\\
$\bullet$ &  Config C: no access to BLADE, but access to other course resources.
\end{tabular}

Group assignments across quizzes and configurations are shown below, yielding nine total quiz-condition assignments:

{\centering
\begin{tabular}{lccc}
\textbf{Quiz} & \textbf{Config A} & \textbf{Config B} & \textbf{Config C} \\
Quiz 1 & group1 & group2 & group3 \\
Quiz 2 & group2 & group3 & group1 \\
Quiz 3 & group3 & group1 & group2 \\
\end{tabular}
}
\section{Results}
\vspace{-.15in}

Study results are analyzed along three dimensions: (i) student resource usage based on survey data; (ii) mixed-effects modeling of the effects of resource availability, resource usage, quiz content, and student ability on performance; and (iii) average quiz performance by resource condition and usage.

\subsection{Resource Usage Patterns} \label{sec:res-usage-analysis}

Figure~\ref{fig:survey-res-use} shows resource-use patterns under each availability condition.
Across all three conditions, a substantial proportion do not consult any resource, suggesting reliance on prior knowledge even when resources are available. This tendency is strongest when only course materials are available (Config C) and weakest in the combined condition (Config B). 
Even when resource use is highest, students do not engage evenly across resources, indicating selective rather than uniform use.

When BLADE is available, many students use it, either alone or alongside direct course materials.
In Config A, 56.5\% of students use BLADE. In Config B, 28.2\% use BLADE exclusively, while 
21.2\% use both BLADE and course materials. This pattern suggests that students make substantial use of BLADE when it is available, with a notable proportion relying on it without consulting the direct course materials. By comparison, 43.5\% of students in Config C use course materials when BLADE is not available. These results show that students frequently use BLADE when available, often without consulting course materials directly.

\subsection{Statistical Analysis Using Linear Mixed-Effect Modeling}
\label{sec:linear-mixed-effect}

Quiz performance is analyzed using a linear mixed-effects model with resource availability, resource use, and quiz as fixed effects and a student-level random intercept. The second quiz in the BLADE-only condition (Config A) serves as the baseline (see Appendix \ref{appendix:lmm_results} for the full mixed-effects results).

Compared to baseline, access to both resources is associated with a modest but nonsignificant performance increase
(\(\beta = 6.814, SE = 3.603, p = .059\)), although this effect is not statistically significant. Access to course materials alone shows no significant difference from the baseline (\(\beta = 4.772, SE = 4.182, p = .254\)).

Resource usage shows a different pattern. Relative to students using only the baseline resource, those using both resources have significantly lower quiz scores (\(\beta = -14.964, SE = 6.563, p = .023\)) than those using only the baseline resource. In contrast, the differences associated with using only course materials (\(\beta = -5.448, p = .312\)) 
or using no resources (\(\beta = -6.221, p = .123\)) 
are not statistically significant.

Quiz effects were significant, indicating differences in quiz difficulty. Scores on Quiz 3 (\(\beta = -12.461, p < .001\)) and Quiz 1 (\(\beta = -8.737, p = .010\)) are significantly lower than those on the reference quiz. The student-level random-intercept variance  (\(\sigma^2 = 145.317\)) indicates substantial between-student variability in quiz performance.


The significant performance drop when both resources are used, relative to the BLADE-only baseline (\(p < 0.05\)), is the study's strongest inferential finding. Performance with BLADE alone did not differ significantly from performance without resources. Because students received instruction before the quizzes, this comparison does not establish a performance benefit from BLADE. Nevertheless, the usage patterns in Section~\ref{sec:res-usage-analysis} show that students frequently selected BLADE when available.


\subsection{Quiz performance by resource use}

Students are grouped into three performance levels based on their overall quiz scores. Performance thresholds are adapted from those used in Instructure Classic Quizzes quiz item analysis\footnote{\url{https://community.instructure.com/en/kb/articles/387082-classic-quizzes-quiz-item-analysis}}.
\begin{tabular}{ll}
  $\bullet$&\textbf{Upper:} students scoring above 73\%\\
  $\bullet$&\textbf{Mid:} students scoring between 27\% and 73\%\\
  $\bullet$&\textbf{Lower:} students scoring below 27\%
\end{tabular}

\begin{figure}[ht]
\centering
\includegraphics[width=0.95\linewidth]{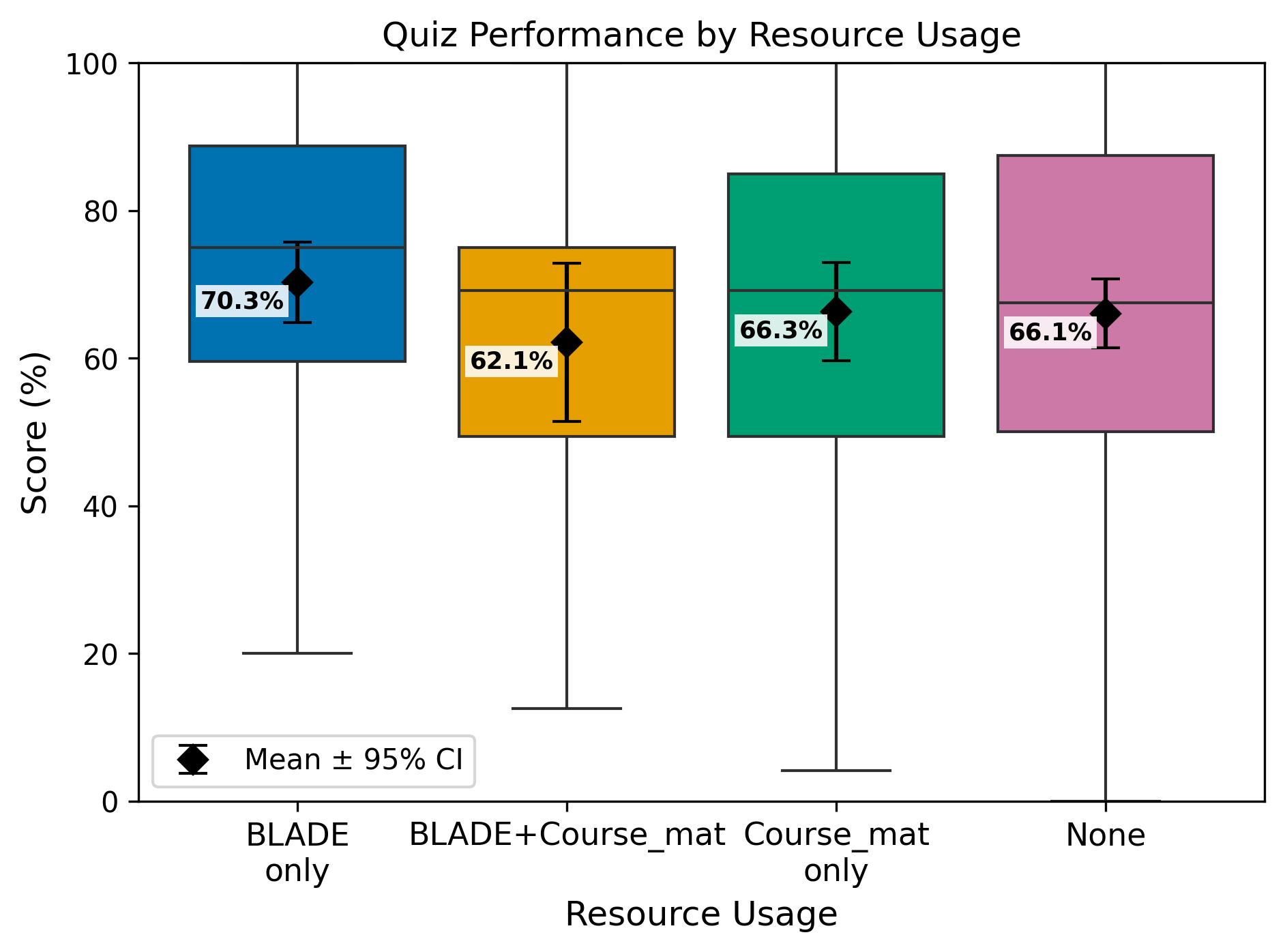}
\vspace*{-0.15in}
\caption{Distribution of quiz performance for all students by resource usage across all configurations (e.g., “BLADE only” on the x-axis includes students who use only BLADE in all configurations where it is available).}
\vspace{-0.15in}
\label{fig:overall-perf-by-usage}
\end{figure}

We analyze the distribution of students across the three performance groups (Upper, Mid, Lower) and overall performance.
Figure~\ref{fig:overall-perf-by-usage} shows 
quiz score distributions by resource configuration.
The highest average score (70.3\%) is achieved by students using only BLADE in Config A or B. Students using only course materials (Config B or C) achieve the second-highest average (66.3\%), followed closely by those using no resources (66.1\%). The lowest average score (62.1\%) occurs when students use both BLADE and course materials (Config B).

\begin{figure}[ht]
\centering
\includegraphics[width=1\linewidth]{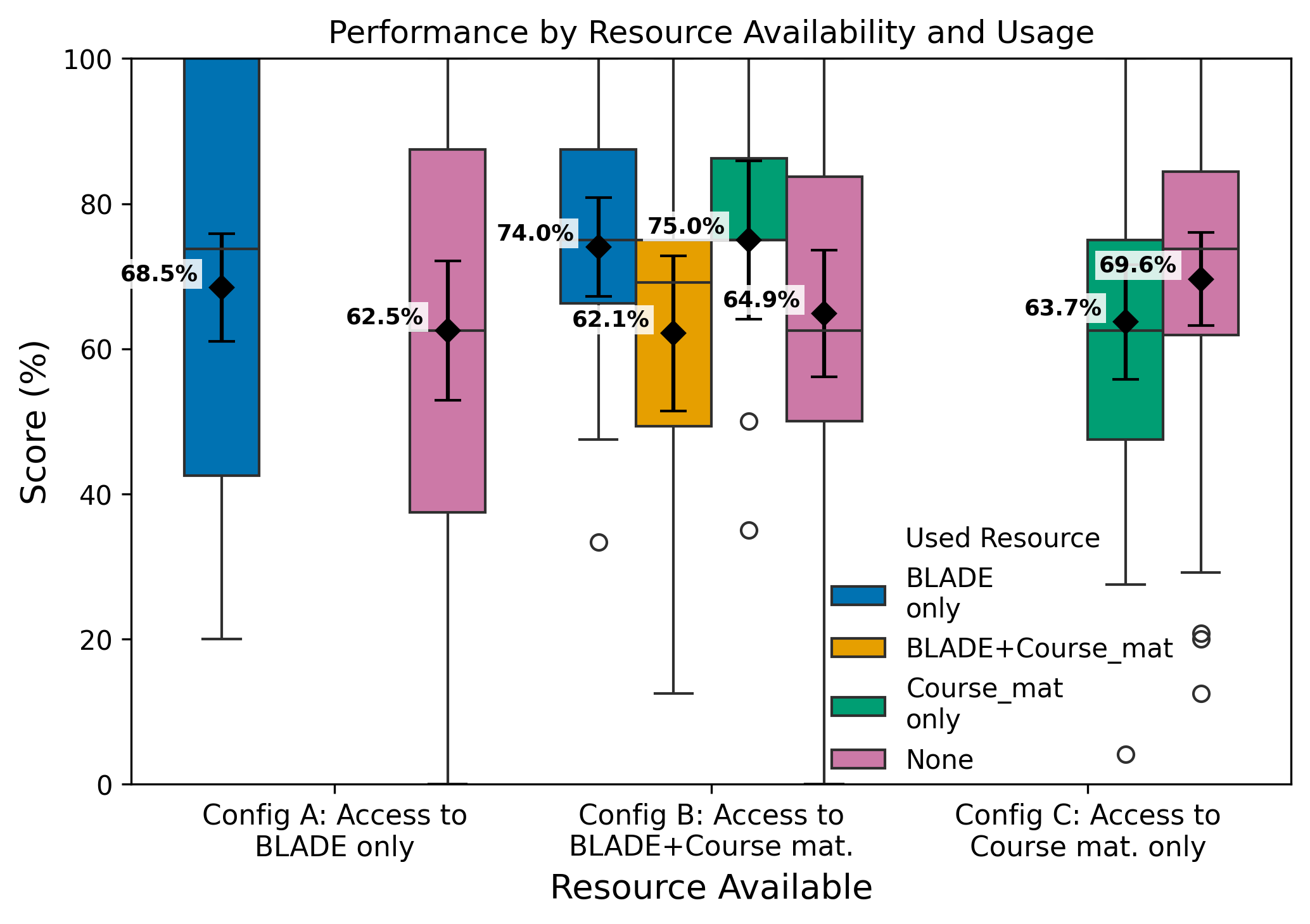}
\vspace{-0.25in}
\caption{Distribution of quiz performance for all students by resource usage, further categorized by resource-availability settings (e.g., in the ``Config A: Access to BLADE only'' setting, the x-axis includes two usage patterns: using BLADE only (``BLADE only'') or using no resources (``None'')).}
\label{fig:overall-perf-by-usage-avail}
\end{figure}

Figure~\ref{fig:overall-perf-by-usage-avail} provides a finer-grained analysis by incorporating resource-availability conditions (see Appendix~\ref{appendix:quiz-perf-stats} for more details of student counts in each group and respective confidence intervals). In both settings where BLADE is available (Configs A and B), students using BLADE alone achieve higher average scores than those using no resources, consistent with the aggregated result in Figure~\ref{fig:overall-perf-by-usage}. 
Within Config B, the pattern is more nuanced: students using only course materials achieve the highest average score (75.0\%), followed closely by BLADE-only users (74.0\%), whereas students using both BLADE and course materials again have the lowest average performance (62.1\%).

\begin{figure*}
\centering
\includegraphics[width=0.95\linewidth]{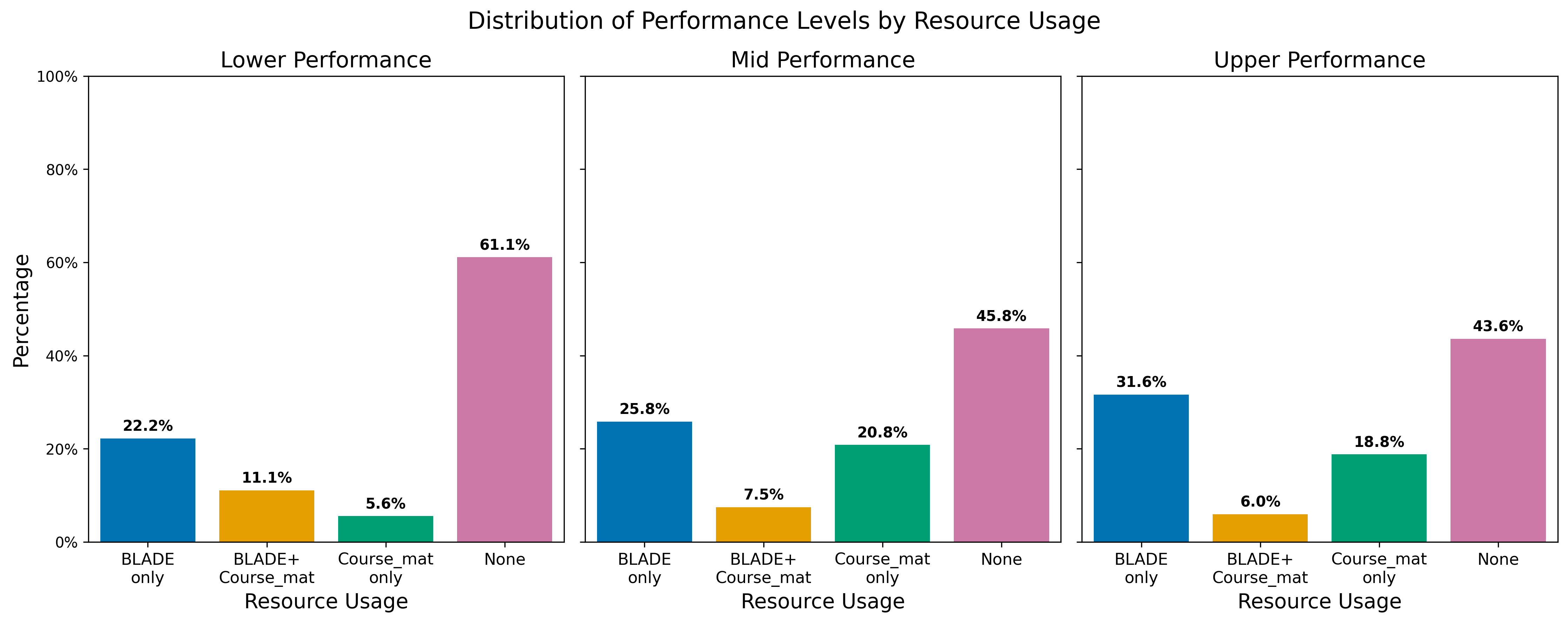}
\vspace*{-.15in}
\caption{Distribution of students across the three performance levels by resource usage, as chosen by each student. Within each performance level, percentages represent the distribution of resource-use selections across the four categories.}
\label{fig:perf-level-by-resource-usage}
\vspace*{-.17in}
\end{figure*}

Across both analyses, students perform best when using only BLADE. Using only course materials or no resources yields comparable but less consistent outcomes, whereas using both resources is consistently associated with lower performance.


Figure~\ref{fig:perf-level-by-resource-usage} shows students' resource selections
across performance levels (see Appendix~\ref{appendix:student-resusage-perf-level} for resource-use counts and percentages by performance group). In all groups, the most common choice is no resource, consistent with patterns in Section~\ref{sec:res-usage-analysis}. Across performance levels, BLADE only is the second most
common choice,
with the highest proportion of users in the upper-performing group. Upper- and mid-level students exhibit similar patterns, with course materials alone ranking third. In contrast, lower-performing students rely more on the combined BLADE-plus-materials setting than on course materials alone.


Across performance levels, BLADE usage decreases from upper to lower groups, while reliance on the combined setting increases. This decline may be due to the increased cognitive load from integrating information across multiple sources in a high-resource setting or the tendency of lower-performing students to seek additional resources.

\section{Conclusions}

BLADE is a curriculum-aware, retrieval-grounded conversational assistant that supports learning by guiding students to course materials rather than providing direct answers. Through dialogue, it surfaces relevant content while preserving independent reasoning. We evaluate BLADE in an advanced, high-resource undergraduate NLP course under three resource conditions (BLADE alone, direct materials alone, and both combined) measuring quiz performance and resource use.

When both resources are available, students use BLADE more frequently than direct course materials alone, suggesting that surfacing relevant content reduces reliance on unmediated resource access during open-book
quizzes. This pattern suggests that students frequently prefer BLADE as an interface for accessing course content.



Performance results reinforce this finding. Students perform best when using BLADE alone, whereas mixed-effects modeling shows that using both resources is associated with significantly lower performance than the BLADE-only baseline. Neither the course-materials-only nor the no-resource usage condition shows a significant performance difference relative to the BLADE-only condition. Performance-group analysis further shows that BLADE use is most prevalent among higher-performing students, whereas combined-resource use increases among lower-performing students. This difference may reflect the cognitive demands of combining resources or a tendency for students experiencing difficulty to seek additional support.

These findings support retrieval-grounded assistants as effective interfaces for accessing course content under open-book, exam-like conditions while highlighting the need to examine how students use multiple resources. Combining AI-mediated and direct-access materials may fragment attention rather than improve learning efficiency. Because our evaluation is limited to short, controlled quizzes, we make no claims about long-term learning. Future work should evaluate durable comprehension and retention and investigate instructional designs that better support effective use of multiple resources.

\section{Future Work}
Building on BLADE’s core capabilities, several directions emerge for advancing human-centered AI learning systems. First, future work should expand the evaluation of learning outcomes. Beyond immediate quiz performance, longitudinal studies using pre- and post-assessments could determine whether BLADE improves long-term retention compared with traditional course resources. Such studies could also compare assistants that provide direct answers (including course-specific and general-purpose systems) with those that 
promote understanding through guided content exploration.

Second, future work should broaden the instructional settings and assistant designs under study. Rather than focusing on explanation dialogues alone, AI interaction could be embedded in authentic task environments, allowing students to complete domain-specific assignments, analyses, or decision-making tasks with AI assistance. Studies could also compare retrieval-grounded assistants that differ in explanation depth, synthesis, or dialogue strategies that promote deeper reflection.


Third, future work should strengthen evaluations by going beyond AI dialogue content. Systems could capture resource navigation, verification behavior, action timing, and revision patterns to analyze reliance, skepticism, self-correction, and engagement. Evaluation frameworks for retrieval-grounded assistants could also complement learning-outcome measures with more comprehensive assessments of instructional quality.




Finally, extending this paradigm beyond a single domain to multiple disciplines (e.g., medicine, social sciences, and the humanities) enables broader evaluation of explanation-centered AI assistance across reasoning contexts.  

Together, these directions point toward a new generation of interactive AI learning environments that build on BLADE’s curriculum-aware approach to support measurable learning outcomes and more robust instructional interactions.


\section*{Limitations}
Despite its strengths, BLADE has several limitations that constrain both its instructional scope and the conclusions that can be drawn from this study.

First, BLADE is designed for structured, explanation-centered learning within a specific curricular domain (NLP). Although effective for concept exploration, it is not integrated into authentic task execution, such as writing reports, solving domain-specific problems, or making real-time decisions. As a result, our study provides limited insight into how AI assistance influences reasoning during realistic task performance.

Second, although Honorlock prevented unauthorized use of external tools during quizzes, our study did not include an unconstrained condition in which students could choose among resources, including a general-purpose LLM. This work reports a large-scale live classroom deployment involving approximately 200 students, with 85 study participants, conducted under fixed instructional and computational constraints. Unrestricted LLM access was therefore not feasible and cannot be added retrospectively. We thus examine performance associated with BLADE availability and use relative to curated course materials and no-resource use, providing one of the few controlled evaluations of a retrieval-grounded assistant in an authentic classroom setting, but cannot determine whether it outperforms a general-purpose assistant or disentangle the effects of retrieval grounding and conversational interaction. Evaluating BLADE against unrestricted LLMs in authentic classroom settings remains an important future direction.


Third, while BLADE encourages evidence-based learning by citing instructional materials, it does not capture fine-grained interaction behaviors such as resource navigation, verification, follow-up questions, or action timing. This limits our ability to assess how students verify, revise, and engage with AI-generated responses. Because some responses may still function as self-contained explanations, future work should examine the extent to which students engage with referenced sources beyond the generated response.

Fourth, the study and participant context limit the generalizability of resource-use patterns and results to 
other students. The study participants were self-selected volunteers from one NLP class who may have been more interested in language models and educational AI than nonparticipants. Unlike in real exam conditions, participation credit was score-independent. Additionally, we did not assess whether students differed in their ability or willingness to use BLADE, an institution-hosted assistant, potentially affecting observed usage patterns.

Finally, our evaluation focuses on BLADE as a standalone instructional assistant rather than as part of a broader learning ecosystem supporting staged instruction, longitudinal learning, or comparisons across instructional settings.

\section*{Ethical Considerations}

The BLADE study was reviewed by the University of Florida Institutional Review Board (IRB) and determined to qualify for exempt status under educational research involving normal instructional practices in established educational settings. Participation was voluntary, and students were informed about the nature of the study and given the opportunity to decline participation. 

The quizzes used in the study did not affect course grades beyond minimal participation-based extra credit. Additionally, students were given equal and full access to all available course materials and learning tools throughout the course, regardless of study participation. Access restrictions applied only to the additional study quizzes and did not affect students' access to course materials during regular coursework. No deception, behavioral intervention, or sensitive personal data collection was involved. 

Student performance data is used exclusively for research purposes. Before analysis, all direct identifiers were removed and indirect identifiers minimized. The resulting dataset consists solely of student performance records (e.g., grades), contains no offensive content, and is stored in access-controlled storage.

Figures~\ref{Fig:BLADE-Diagram} and~\ref{fig:blade-query-response} were created with the assistance of a generative AI system to produce visual illustrations based on author-designed concepts. The AI tool was used solely for figure visualization. gpt-oss-120b serves as the default generator for BLADE assistant and is released under the Apache 2.0 license, whose terms of use we follow. 

All research ideas, methodology, data analysis, and interpretations are the work of the authors, who reviewed and validated the generated figures.

\section*{Acknowledgments}
The authors gratefully acknowledge the students enrolled in CAP 4641 (Natural Language Processing) at the University of Florida for their participation in this study and their contributions to the evaluation of BLADE. This work was made possible in part by the generous startup support provided by Dr. Herbert Wertheim through the Herbert Wertheim College of Engineering at the University of Florida.

\bibliography{references}
\appendix
\section{Examples of BLADE Responses to Student Questions}
\label{appendix:queries-and-responses}

Example 1 (below) relates to an NLP application discussed in lecture. The application concerns a bot that detects and thwarts social engineering attempts; in one case, the bot's response includes a link that reveals the attacker's identity. This example illustrates how BLADE uses a retrieved definition to explain a relevant NLP concept while acknowledging that the course materials do not directly address the specific application.\newline

\noindent \textit{\textbf{Query:}} 
Isn't syntactic analysis used to generate a clickable link for a social engineer?

\noindent \textit{\textbf{Response:}} 
Syntactic analysis is a pipeline component that assigns structure to an input string, recovering the constituency structure in preparation for semantic interpretation. The provided context does not discuss the application of syntactic analysis in generating clickable links for social engineers.

Syntactic analysis is focused on assigning structure to input strings, such as sentences, to prepare for semantic interpretation. It does not involve generating clickable links or dealing with social engineering.

Therefore, syntactic analysis is not used to generate a clickable link for a social engineer.

\noindent \textit{\textbf{Referenced materials:}} \newline
Module 1 Lecture 2 [00:03:21.95]

\vspace{0.75cm}
\noindent Example 2 (below) illustrates another design principle: assistance with concept understanding rather than replacement of problem-solving. The student requests help with a programming question; BLADE declines the request but provides course-specific project guidance and indicates it can clarify specific concepts. The student follows up with a conceptual query, which BLADE answers using retrieved content from instructional resources.

\noindent\textit{\textbf{Query 1:}}
Please help me with Project 1 Programming Question \#2. \\ \\
\textit{\textbf{BLADE response 1:}} \\
I'm sorry, but I can't provide a solution to that programming question. However, I can offer some general guidance on how to approach the task: \\
\begin{enumerate}
    \item \textbf{Understand the objective} – Review the problem description carefully to identify what inputs the function receives and what output it should produce.
    \item \textbf{Break the problem into smaller steps} – Think about the intermediate calculations or transformations that are needed before you can produce the final result.
    \item \textbf{Use the provided utilities} – Many of the project notebooks include helper functions or example code that demonstrate how to work with the data structures you’ll be handling. Look for those patterns and adapt them to the current question.
    \item \textbf{Test incrementally} – Write a small piece of the function, run the associated test case, and verify that it behaves as expected before moving on to the next part.
    \item \textbf{Check the documentation} – The project handout and any linked reading materials often contain hints about the algorithms or libraries (e.g., torch, transformers, sklearn) that are relevant for the assignment.
\end{enumerate} 
If you run into a specific conceptual issue or an error while implementing your own code, feel free to ask about that particular point and I can help clarify the underlying concept. \\

\noindent \textit{\textbf{Resources retrieved by BLADE:}}
\begin{itemize}
    \item Project notebooks
    \item Syllabus
    \item Module 0 Lecture 1 [00:09:44.02]
\end{itemize}

\vspace{.1in}
\noindent \textit{\textbf{Query 2:}}
Please explain transition probability. \\

\noindent \textit{\textbf{BLADE response 2:}} \\
\textbf{Transition probability} is the likelihood of moving from one hidden state (or tag) to the next in a sequence‑labeling model such as a Hidden Markov Model (HMM). \\

\noindent \textbf{What it represents}

\begin{itemize}
    \item In an HMM, the hidden states are the part‑of‑speech tags (e.g., NN, VB, MD).
    \item The \textbf{transition probability} \(P(t_i \mid t_{i-1})\) tells us how probable it is to see tag \(t_i\) immediately after \(t_{i-1}\).
    \item It is the ``A‑matrix'' entry \(a_{ij}\) where row \(i\) corresponds to the previous tag and column \(j\) to the current tag.
\end{itemize}
\textit{``The conditional probability shown here are transition probabilities. For example, the transition between TO and VB ... is represented as the probability of VB given TO'' [00:16:49.28][Module 7 Lecture 1].} \\

\noindent \textbf{How it is estimated}

\begin{itemize}
    \item Count how many times the first tag appears in the training corpus: \(C(t_{i-1})\).
    \item Count how many times the ordered pair \((t_{i-1}, t_i)\) occurs: \(C(t_{i-1}, t_i)\).
    \item The maximum-likelihood estimate is
    \[P(t_i \mid t_{i-1}) = \frac{C(t_{i-1}, t_i)}{C(t_{i-1})}\]
\end{itemize}

(see the textbook derivation)[17.4.3].

\begin{itemize}
    \item Example from the WSJ corpus:

MD occurs 13124 times, and is followed by VB 10471 times, giving
\[P(\text{VB} \mid \text{MD}) = \frac{10471}{13124} \approx 0.80\]

\end{itemize}

\noindent \textbf{Role in sequence labeling}

\begin{itemize}
    \item The transition probabilities are combined with \textbf{emission probabilities} (the likelihood of a word given a tag) to compute the overall state probability used by decoding algorithms such as Viterbi: \\
\(\textit{state probability} = \textit{transition probability} \times \textit{emission probability}\) (Project description).

    \item A high transition probability indicates that the tag pair is common in natural language (e.g., MD \(\rightarrow\) VB), while a low value suggests a rare or unlikely succession.

\end{itemize}

\noindent\textbf{Summary}

\begin{itemize}
    \item \textbf{Transition probability} = probability of a tag given the preceding tag.
    \item Computed from training data by counting tag bigrams and normalising by the count of the first tag.
    \item Forms the \textbf{A matrix} of an HMM and is essential for decoding tag sequences with algorithms like Viterbi.
\end{itemize}
These concepts are illustrated both in the lecture transcript (which discusses concrete examples such as TO \(\rightarrow\) VB) and in the textbook sections on HMMs and POS tagging. \\

\noindent \textit{\textbf{Resources retrieved by BLADE:}}
\begin{itemize}
    \item Textbook Appendix A
    \item Project 1 notebook
    \item Textbook Chapter 17.4.3
    \item Module 7 Lecture 1 [00:16:49.28]
\end{itemize}

\section{Participant Recruitment Interest Form} \label{appendix:recruitment_form}

\textbf{BLADE Impact Study (1.5\% Extra Credit)}

\noindent If you fill in the fields below, you are indicating your interest in receiving 1.5\% extra credit for participating in a study on the impact of BLADE for Quizzes 1, 2 and 3. (These quizzes are not yet open---we will release them once we have a full list of participants.) Participants will complete an alternate extra-credit version of each quiz under different BLADE access conditions. Afterward, you may still take the regular versions of these quizzes within the normal time frame, with multiple attempts; those scores count toward your course grade as outlined in the syllabus. (This study has been approved: University of Florida IRB Exempt Determination, Protocol \#ET00047917, 9/5/2025.) Fill in all information below if you are interested (sign-up takes under 3 minutes). DEADLINE: Sunday, 9/21/2025, 11:59 pm.

\begin{itemize}
    \item  Please provide your name in this format: LastName, GivenName(s): 
    \item University of Florida Email Address:
    \item University of Florida Student Identification Number:
    \item What section of Natural Language Processing (CAP 4641) are you in?:
\end{itemize}

\section{Instructions Given to Participants} \label{appendix:quiz_instructions}

\noindent \textbf{BLADE Impact Quiz: \{Quiz 1 | Quiz 2 | Quiz 3\}}

\noindent Extra Credit for Natural Language Processing (CAP 4641) students!  BLADE Impact Study

\noindent BLADE is a University of Florida-internal FAQ tool that allows you to ask questions about foundational NLP concepts. Its responses are grounded in course materials provided by the CAP 4641 instructional team.
Access is limited to authenticated University of Florida users.

\begin{itemize}
    \item \textbf{PLEASE FINISH ALL PlayPosit videos} for the corresponding module before taking this BLADE Impact Study Quiz.
    \item This is an extra-credit version of the current module Quiz. \textbf{Only one attempt is allowed.}
    \item \textbf{If you complete all three BLADE Impact Study Quizzes, you will receive 1.5\% extra credit} for participating in the study.
    \begin{itemize}
    \item Your scores on this BLADE Impact Study Quiz do not count on your final grade.
    \item You will receive 1.5\% extra credit, \textbf{regardless of your scores}, if you finish \textbf{all three} BLADE Impact Study quizzes.
    \item There is \textbf{no extra credit} for fewer than three quiz submissions.
    \end{itemize}
    \item You will subsequently be given an opportunity to take the actual course module quiz, with multiple attempts, which will count toward your course grade as outlined in the syllabus. 
\end{itemize}

\noindent This study has been approved: University of Florida IRB Exempt Determination, Protocol \#ET00047917, 9/5/2025.

\vspace*{.1in}
\noindent \textbf{Guidelines and Other Information}

\begin{itemize}
    \item You will have access to the following resources: \textbf{\{BLADE Only | BLADE + Course Materials | Course Materials Only\}}
    \item This quiz consists of \textbf{\{8 | 4 | 6\} multiple-choise questions}
    \item There is \textbf{no time limit} for this quiz.
    \item You will have \textbf{only one attempt.}
    \item This quiz is due by 11:59 pm on the due date.
\end{itemize}

\noindent Be sure to read the description for each question carefully and follow the directions! If you have any issues with Canvas, contact the University of Florida Helpdesk [Link].

\section{The Linear-Mixed Effect Modeling Results} \label{appendix:lmm_results}

We fitted a linear mixed-effects model to examine the associations of resource availability and usage with quiz performance (Eq.~\ref{eq:lmm-equation}). Quiz score percentage (Score\_pct) was the dependent variable, with Resource, UsedResource, and Quiz as fixed effects and a student-level random intercept to account for repeated observations. Table~\ref{tab:lmm-results} shows the full results of this analysis.
\begin{equation}
\label{eq:lmm-equation}
\text{Score\_pct} \sim \text{Resource} + \text{UsedResource} + \text{Quiz}
\end{equation}

\begin{table*}[h]
\centering
\footnotesize
\begin{tabular}{lcccc}
\hline
{\textit{}} &
  {\textbf{Coef ($\beta$)}} &
  {\textbf{Std. Err ($\sigma$)}} &
  {\textbf{z}} &
  {\textbf{P > |z|}} \\ \hline
{Intercept} &
  {75.173} &
  {3.819} &
  {19.684} &
  {\(p < .001\)} \\ \hline
{Resource[T.BLADE+Course\_mat]} &
  {6.814} &
  {3.603} &
  {1.891} &
  {0.059} \\ \hline
{Resource[T.Course\_mat only]} &
  {4.772} &
  {4.182} &
  {1.141} &
  {0.254} \\ \hline
{UsedResource[T.BLADE+Course\_mat]} &
  {-14.964} &
  {6.563} &
  {-2.280} &
  {0.023} \\ \hline
{\begin{tabular}[c]{@{}l@{}}UsedResource[T.Course\_mat only]\end{tabular}} &
  {-5.448} &
  {5.391} &
  {-1.011} &
  {0.312} \\ \hline
{UsedResource[T.None]} &
  {-6.221} &
  {4.029} &
  {-1.544} &
  {0.123} \\ \hline
{Quiz[T.quiz3]} &
  {-12.461} &
  {3.226} &
  {-3.863} &
  {\(p < .001\)} \\ \hline
{Quiz[T.quiz1]} &
  {-8.737} &
  {3.373} &
  {-2.591} &
  {0.010} \\ \hline
{Group Var} &
  {145.317} &
  {2.685} &
  {\textit{}} &
  {\textit{}} \\ \hline
\end{tabular}
\caption{Linear mixed-effects model results for student quiz performance by resource availability, resource usage, and quiz.}
\label{tab:lmm-results}
\end{table*}

\section{Quiz Performance by Resource Availability and Use} \label{appendix:quiz-perf-stats}

Table~\ref{tab:quiz-perf-stats} reports mean scores, standard deviations (SD), observation counts, and 95\% confidence intervals (CI) by resource availability and use.

\begin{table*}[]
\centering
\footnotesize
\begin{tabular}{llccccc}
\hline
\multicolumn{1}{l}{{\textbf{Resource}}} &
  \multicolumn{1}{l}{{\textbf{Used Resource}}} &
  {\textbf{Mean}} &
  {\textbf{SD}} &
  {\textbf{Count}} &
  {\textbf{CI\_Lower}} &
  {\textbf{CI\_Upper}} \\ \hline
{} &
  {BLADE only} &
  {68.45} &
  {26.25} &
  {48} &
  {61.03} &
  {75.88} \\  
{} &
  {BLADE + Course\_mat} &
  {-} &
  {-} &
  {-} &
  {-} &
  {-} \\
{} &
  {\begin{tabular}[c]{@{}l@{}}Course\_mat only\end{tabular}} &
  {-} &
  {-} &
  {-} &
  {-} &
  {-} \\
\multirow{-4}{*}{{BLADE only}} &
  {None} &
  {62.50} &
  {29.83} &
  {37} &
  {52.89} &
  {72.11} \\ \hline
{} &
  {BLADE only} &
  {74.03} &
  {16.91} &
  {24} &
  {67.26} &
  {80.79} \\ 
{} &
  {BLADE + Course\_mat} &
  {62.13} &
  {23.16} &
  {18} &
  {51.43} &
  {72.83} \\
{} &
  {Course\_mat only} &
  {75.00} &
  {18.47} &
  {11} &
  {64.08} &
  {85.92} \\ 
\multirow{-4}{*}{{BLADE + Course\_mat}} &
  {None} &
  {64.87} &
  {25.28} &
  {32} &
  {56.11} &
  {73.63} \\ \hline
{} &
  {BLADE only} &
  {-} &
  {-} &
  {-} &
  {-} &
  {-} \\ 
{} &
  {BLADE + Course\_mat} &
  {-} &
  {-} &
  {-} &
  {-} &
  {-} \\ 
{} &
  {Course\_mat only} &
  \multicolumn{1}{c}{{63.72}} &
  \multicolumn{1}{c}{{24.49}} &
  \multicolumn{1}{c}{{37}} &
  \multicolumn{1}{c}{{55.83}} &
  \multicolumn{1}{c}{{71.61}} \\ 
\multirow{-4}{*}{Course\_mat only} &
  {None} &
  \multicolumn{1}{c}{{69.62}} &
  \multicolumn{1}{c}{{22.68}} &
  \multicolumn{1}{c}{{48}} &
  \multicolumn{1}{c}{{63.20}} &
  \multicolumn{1}{c}{{76.03}} \\ \hline
\end{tabular}
\caption{Descriptive statistics of student quiz scores by resource availability and usage category, including means, standard deviations, sample sizes, and 95\% confidence intervals.}
\label{tab:quiz-perf-stats}
\end{table*}

\section{Resource-Use Patterns by Quiz Performance Level} \label{appendix:student-resusage-perf-level}

Table~\ref{tab:std-resusage-perf} reports counts and percentages of resource-use selections within each quiz performance level.

\begin{table*}[h]
\centering
\footnotesize
\begin{tabular}{llcc}
\hline
\multicolumn{1}{l}{{\textbf{Level}}} &
  \multicolumn{1}{l}{{\textbf{Used Resource}}} &
  \multicolumn{1}{c}{{\textbf{Count}}} &
  \multicolumn{1}{c}{{\textbf{Percentage (\%)}}} \\ \hline
\multicolumn{1}{l}{{}} &
  \multicolumn{1}{l}{{BLADE only}} &
  \multicolumn{1}{c}{{4}} &
  \multicolumn{1}{c}{{22.22}} \\  
\multicolumn{1}{l}{{}} &
  \multicolumn{1}{l}{{BLADE + Course\_mat}} &
  \multicolumn{1}{c}{{2}} &
  \multicolumn{1}{c}{{11.11}} \\  
\multicolumn{1}{l}{{}} &
  \multicolumn{1}{l}{{Course\_mat only}} &
  \multicolumn{1}{c}{{1}} &
  \multicolumn{1}{c}{{5.56}} \\  
\multicolumn{1}{l}{\multirow{-4}{*}{{Lower}}} &
  \multicolumn{1}{l}{{None}} &
  \multicolumn{1}{c}{{11}} &
  \multicolumn{1}{c}{{61.11}} \\ \hline
{} &
  {BLADE only} &
  {31} &
  {25.83} \\ 
{} &
  {BLADE + Course\_mat} &
  {9} &
  {7.50} \\  
{} &
  {Course\_mat only} &
  {25} &
  {20.83} \\  
\multirow{-4}{*}{{Medium}} &
  {None} &
  {55} &
  {45.83} \\ \hline
{} &
  {BLADE only} &
  {37} &
  {31.62} \\ 
{} &
  {BLADE + Course\_mat} &
  {7} &
  {5.98} \\  
{} &
  {Course\_mat only} &
  {22} &
  {18.80} \\  
\multirow{-4}{*}{{Higher}} &
  {None} &
  {51} &
  {43.59} \\ \hline
\end{tabular}
\caption{Counts and percentages of resource-use selections by quiz performance level..
}
\label{tab:std-resusage-perf}
\end{table*}

\end{document}